\documentclass[fleqn,usenatbib]{mnras}

\usepackage{graphicx}
\usepackage{amsmath,amssymb}
\usepackage{caption}
\usepackage{subcaption}
\usepackage{txfonts}
\usepackage{amsmath}
\usepackage{geometry}
\usepackage[T1]{fontenc}

\DeclareRobustCommand{\VAN}[3]{#2}
\let\VANthebibliography\thebibliography
\def\thebibliography{\DeclareRobustCommand{\VAN}[3]{##3}\VANthebibliography}

\usepackage{graphicx}	% Including figure files
\usepackage{amsmath}	% Advanced maths commands

\title[Afterglow study of GRB~250129A]{Photometric and late-time spectropolarimetric observations of GRB~250129A afterglow} 
\author[A. Ghosh et al.]{A. Ghosh$^1$,\thanks{E-mail: ghosh.ankur1994@gmail.com (AG)}
S. Razzaque$^{1,2,3}$\thanks{E-mail: srazzaque@uj.ac.za},
J. Barnard$^{4}$,
J. C. Joshi$^{5,1}$,
R. Gupta$^{7}$,
D. A. H. Buckley$^{7,8}$,
B. van Soelen$^{4}$,
\newauthor
N. Dukiya$^{5,9}$,
A. Gupta$^{5,10}$,
A. S. Moskvitin$^{11}$,
J. Cooper$^{4}$,
S. Chandra$^{12}$,
K. M. Jayasurya$^{13}$,
K. Misra$^{5}$,
\newauthor
N. Rawat$^{8}$,
L. Resmi$^{12}$,
O. I. Spiridonova$^{11}$,
R. I. Uklein$^{11}$.
\\
$^1$Centre for Astro-Particle Physics (CAPP) and Department of Physics,
University of Johannesburg, PO Box 524, Auckland Park 2006,
Johannesburg, South Africa\\
$^2$Department of Physics, The George Washington University,
Washington, DC 20052, USA\\
$^3$National Institute for Theoretical and Computational Sciences
(NITheCS), Private Bag X1, Matieland, South Africa\\
$^{4}$Department of Physics, University of the Free State, 205 Nelson
Mandela Dr., Bloemfontein, 9300, South Africa\\
$^{5}$Aryabhatta Research Institute of Observational Sciences, Manora Peak,
Nainital, 263001, India\\
$^{6}$Astrophysics Science Division, NASA Goddard Space Flight Center,
Mail Code 661, Greenbelt, MD 20771, USA\\ 
$^{7}$AAA Department of Astronomy, University of Cape Town, Private Bag
X3, Rondebosch 7701, South Africa\\
$^{8}$AAB South African Astronomical Observatory, PO Box 9, 7935
Observatory, Cape Town, South Africa\\
$^{9}$Department of Applied Physics, Mahatma Jyotiba Phule Rohilkhand
University, Bareilly, 243006, India\\
$^{10}$Department of Physics, Indian Institute of Technology Roorkee,
Roorkee, Nainital, 247667, India\\ 
$^{11}$ Special Astrophysical Observatory, Russian Academy of Sciences,
Nizhnii Arkhyz, 369167, Russia\\
$^{12}$Indian Institute of Space Science and Technology, Trivandrum, 695547,
Kerala, India\\
$^{13}$Space Astronomy Group, ISITE Campus, U. R. Rao Satellite Centre, Bangalore, 560037, India.\\
}
\date{Accepted XXX. Received YYY; in original form ZZZ}

\pubyear{\the\year{}}

\begin{document}
\label{firstpage}
\pagerange{\pageref{firstpage}--\pageref{lastpage}}
\maketitle
% Abstract of the paper
\begin{abstract}
Gamma-Ray Burst (GRB) afterglows arise from the interaction of relativistic ejecta with the circumburst medium and are observed across the electromagnetic spectrum. Afterglow polarisation is expected at early and late afterglow phases depending on the presence of reverse shocks (RS) and the observer’s viewing geometry relative to the jet. Polarimetric observations of GRB afterglows serve as a unique diagnostic tool to investigate the geometry and structure of magnetic fields in the emitting region, which cannot be directly inferred from photometric or spectroscopic data alone. We report late-time ($\sim$ 19 hours post-burst) spectropolarimetric observations of GRB~250129A using the Southern African Large Telescope (SALT). The data reveal a hint of linear polarisation, but rotation in polarisation angle across wavelengths is not observed. Polarisation is usually expected during the early afterglow ($\lesssim$ 100 s) when the RS dominates. However, multi-wavelength modelling of GRB~250129A data shows no indication of RS contribution during the late epoch. Afterglow model incorporating both the forward shock (FS) and RS confirms that RS component fades rapidly after $\sim$100 s. The  afterglow emission is best explained by an off-axis viewing geometry of a jet with Gaussian core and wing evolving in a uniform density environment. GRB~250129A thus provides a rare observational evidence linking late-time polarisation to geometric and jet-structure effects.
\end{abstract}

% Select between one and six entries from the list of approved keywords.
% Don't make up new ones.

%%%%%%%%%%%%%%%%%%%%%%%%%%%%%%%%%%%%%%%%%%%%%%%%%%

%%%%%%%%%%%%%%%%% BODY OF PAPER %%%%%%%%%%%%%%%%%%

\section{Introduction}

Gamma-ray bursts (GRBs) are the most luminous electromagnetic explosions in the Universe, releasing enormous isotropic energy (between $10^{48} - 10^{55}$ ergs within a short timescale. Despite resulting from the core collapse of a massive star or merger of binary neutron stars (BNS), the central engine of GRBs (a black hole or a millisecond magnetar) accretes the infalling matter and launches a bipolar ultra-relativistic jet that expands in the ambient medium. The interaction between the jet and surrounding medium produces long-term external shock emission in radio to $\gamma$-rays, broadly named as afterglow emission \citep{1997ApJ...476..232M, 1998ApJ...497L..17S, 1999PhR...314..575P}. These collision driven shocks are bidirectional in nature: one expands towards unshocked surrounding medium called forward shock (FS) and the later one propagates backward towards shocked ejecta known as reverse shock (RS) \citep{1999ApJ...519L..17S, 2000ApJ...545..807K, 2005ApJ...628..315Z}. Afterglow studies provide a unique diagnostic of relativistic outflow physics, enabling constraints on the jet and surrounding environmental properties, and microphysical parameters governing particle acceleration and magnetic field amplification. Despite decades of observational and theoretical studies, fundamental questions yet remain unresolved such as, jet structure and the configuration of magnetic fields within the emitting region \citep{2004RvMP...76.1143P, 2015PhR...561....1K}. 

%importance of polarisation in afterglow study and early time polarisation detection
Afterglow studies heavily rely on photometry and broadband modelling, which provide indirect constraints on jet structure and configuration, the microphysics of shock generated electrons. However, these approaches are largely insensitive to geometric and magnetic field properties, since flux measurements encode only isotropic intensity. Polarisation, in contrast, is a powerful tool for the direct diagnostics of beaming and orientation information of the jet, jet structure, and magnetic field of the outflows. The detailed afterglow polarisation of a few GRBs in the past has been reviewed by \citet{2016Covino} in detail. Synchrotron emission, the mechanism mostly influence the afterglow radiation, is intrinsically weakly polarised \citep{1997ApJ...476..232M}. For a spherical outflow with a random magnetic field viewed from on-axis, the net polarisation averages out to zero. In contrast, the early afterglow ($\sim$ up to few hundred seconds), mostly dominated by the RS emission, shows very high linear polarisation \citep{2013Natur.504..119M, 2017Natur.547..425T, 2024NatAs...8..134A}. The early afterglow emission which is Poynting flux dominated, is expected to be highly polarised due to the presence the magnetic field in the ejecta inherited from the central engine. 

%late-time polarisation detection
The late time GRB afterglow is mostly dominated by the forward shock. In this regime, the PD depends sensitively on the observer’s viewing angle relative to the jet opening angle as well as the jet’s internal structure. A very low level (a few \%) of polarisation is expected in the FS regime, while in off-axis cases the PD could in principle reach up to 50\% at late-time. Although, such high level of polarisation has never been observed till date \citep{2022A&A...666A.179B}. Strong polarisation was mostly detected at very early phase of afterglow \citep{2009Natur.462..767S, 2012ApJ...752L...6U, 2013Natur.504..119M}. Spectropolarimetry at optical wavelengths may provide some additional information about the presence of synchrotron break frequencies at optical wavelengths and allow correction for dust-induced polarisation in both the Milky Way and the GRB host galaxy. Detection of late-time polarisation in GRB afterglows through spectropolarimetric observation is very rare. To date, late-time polarisation observation through spectropolarimetry was reported only in four GRBs (GRB~020813 \citealt{2003ApJ...584L..47B}, GRB~030329A \citealt{2003Natur.426..157G}, GRB~191221B \citealt{2021MNRAS.506.4621B}, GRB~080928 \citealt{2022A&A...666A.179B}.) High PD was quoted only for GRB~030329A at late-phase but during the associated supernova emission.

Our study presents a comprehensive photometric and spectropolarimetric study of GRB~250129A afterglow. In Section \ref{ch2:GRB}, we provide a brief overview of GRB~250129A properties. Sections \ref{ch3:data} and \ref{ch4:specpol} includes the photometry and spectropolarimetry data observation and analysis. In Section~\ref{ch5:model}, we present theoretical interpretations of the GRB afterglow and observed polarisation. Section~\ref{ch6:res} represents the afterglow and polarimetric results and in Section~\ref{ch7:conclu}, we summarise the key implications of our results.

\section{GRB~250129A}
\label{ch2:GRB}

GRB~250129A was first triggered and reported by the Neil Gehrels \textit{Swift} Observatory (\textit{Swift},  \citealt{2004ApJ...611.1005G} -  burst alert telescope (BAT,  \citealt{2005Barthelmy}) on 29 January 2025 at 04:45:09 UT \citep{Beardmore2025}. The BAT light curve of GRB~250129A shows a complex multi-peak emission extended up to $\sim$ 262 s since the trigger. The \textit{Swift} -  X-ray telescope (XRT,  \citealt{2005Burrows}) slewed to the burst position $\sim$ 156 s since the trigger to locate the burst precisely. The first phase of XRT data was observed during the prompt emission using the windowed timing (WT) mode, while the remaining data was observed in photon counting (PC) mode from $\sim$ 1000 - 46000 s since the trigger. The \textit{Swift} - Ultraviolet and Optical Telescope (UVOT,  \citealt{Roming2005}) observed GRB~250129A $\sim$ 1184 s since the BAT trigger with an exposure time of 150 s in the white filter. A bright afterglow in X-ray and optical was detected by XRT and UVOT, respectively \citep{Beardmore2025}. GRB~250129A was triggered by only $\gamma$-ray satellite \textit{Konus-Wind}. The fluence measured in the 10 keV - 10 MeV band is $6.71 \pm 0.14 \times 10^{-6}$ erg/$ \rm cm^2$ \citep{2025Frederiks}. Using the measured redshift of 2.151 by \citet{2025Schneider} and standard cosmological parameters of $H_0$ = $69.6~\rm km~s^{-1}~Mpc^{-1}$, $\Omega_m$ = $0.286$, and $\Omega_\Lambda$ = $0.714$ by \citet{2014ApJ...794..135B}, we calculated the isotropic energy $E_{\rm iso}$ = $(7.7 \pm 0.2) \times 10^{52}$ erg and isotropic peak luminosity $L_{\rm iso}$ = $(5.3 \pm 0.7) \times 10^{51}$ erg/s respectively. There is no hint of detection of polarisation in the high energy regime. 

The first optical afterglow was observed by MASTER-OAFA robotic telescope (Global MASTER-Net,\footnote{\url{http://observ.pereplet.ru}} \citep{2010Lipunov}) at 2025-01-29 04:50:05 UT ($\sim$ 296 s after the burst trigger). A bright afterglow was detected in r filter with 17.5 mag. We searched for the optical transient (OT) associated with GRB~250129A at the \textit{Swift}-UVOT position using the 0.4-m SCICAM QHY600 at the Las Cumbres Observatory Global Telescope (LCOGT) node located at Teide Observatory, Tenerife \citep{2025Ghosh}. Because of the bright early afterglow emission, spectroscopic observations were carried out using different telescopes (VLT, NOT, OHP/T193) globally who reported the redshift z = 2.151 \citep{2025Schneider, 2025Izzo, 2025Schneiderb}. The Space Variable Objects Monitor (SVOM)/  Chinese Ground Follow-up Telescope (C-GFT) also detected the OT $\sim$ 37.11 hr after the trigger using g, r, i bands \citep{2025GCN.39124....1W}. Due to its slow decaying nature, multiple telescopes observed GRB~250129A for several epochs and reported in GCN circulars. We continued the follow-up observations of GRB~250129A with the 1-m Sinistro telescopes of LCO, the SAO-RAS Zeiss-1000 telescope, and the Devasthal optical telescope (DOT). \textit{Swift} X-ray telecope (Swift-XRT) light curve and time-sliced spectra at three different epochs (0.051, 0.069, 040 days post-burst) were extracted from \textit{Swift} XRT repository.\footnote{ \url{https://www.swift.ac.uk/xrt_curves/01285812/}} The observed count rates were converted into a flux density light curve at 10 keV using the standard formulations.

\section{Optical photometry}
\label{ch3:data}

   \begin{figure}
   \centering
   \includegraphics[width=\columnwidth]{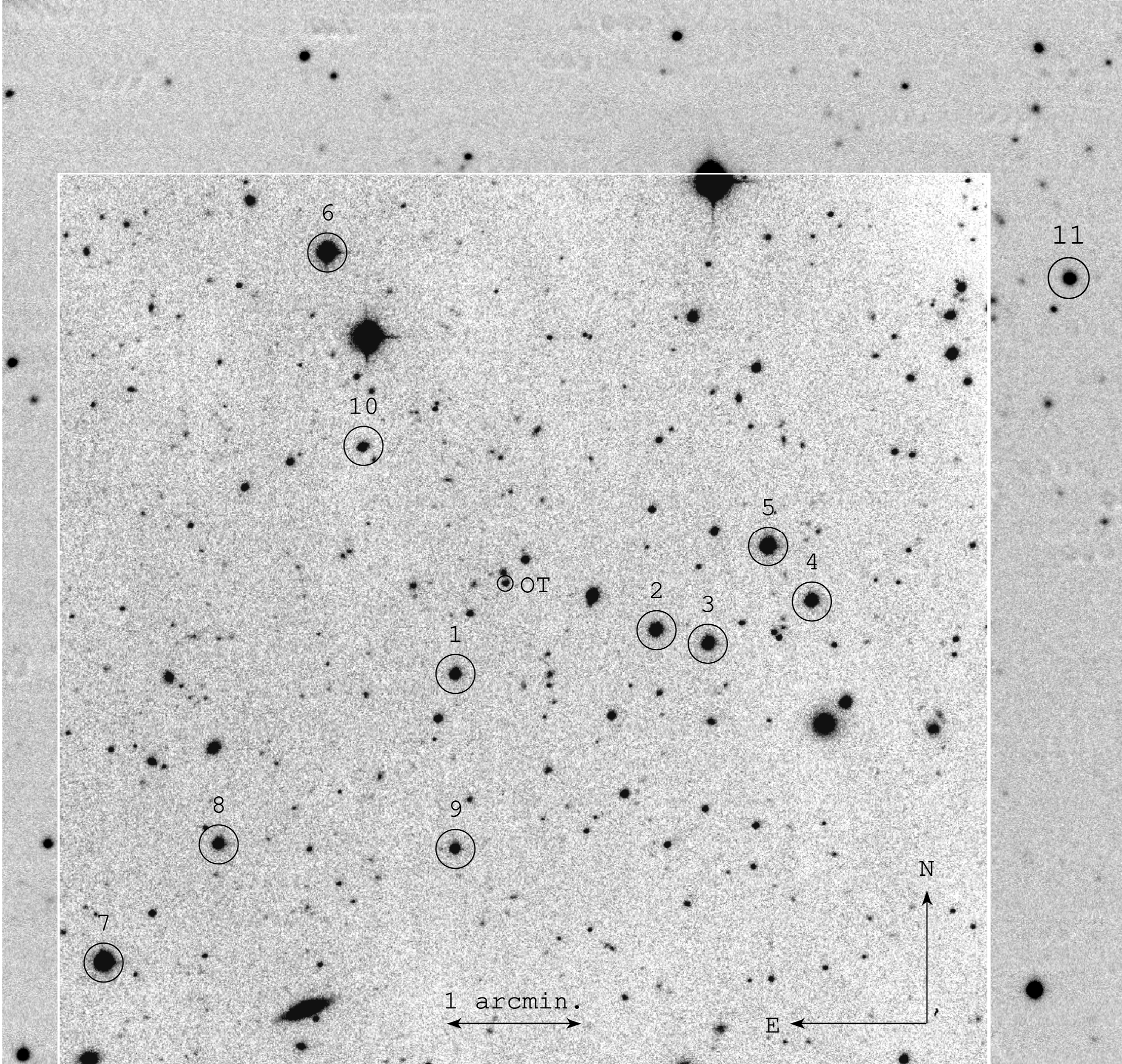}
      \caption{Finding chart of GRB 250129A field (the CCD-photometer FoV placed ontop of central square of MAGIC FoV). Positions of OT and used standard stars are marked with the circles, their magnitudes are listed in the Table \ref{standards}}
         \label{Fig:chart}
   \end{figure}

\subsection{LCO photometry of GRB~250129A}
\label{LCO}

We observed the field of GRB~250129A with different 0.4-m and 1-m LCO telescopes worldwide as a part of our proposal ``Optical follow-up of gamma-ray transients'' (proposal id: SAAO2024B-005). Every 0.4-m and 1-m LCO telescope has an identical configuration and is used for imaging purposes. The 1-m Sinistro telescopes are equipped with 4k $\times$ 4k CCD imagers cover the field of view (FOV) $26.6^{\prime} \times 26.6^{\prime}$. In contrast, the 0.4-m SCICAM QHY600 is equipped with 9576 x 6388 pixel CCD (FOV: 1.9 x 1.2 degrees) but we only used the FOV of $30^{\prime} \times 30^{\prime}$ for our observation \citep{2013Brown}. Our observation started from 1.17 hours and continued until $\sim$ 5 days since the GRB trigger using B, V, and r filters. The exposure time spanning from 330 - 1200 s was chosen depending on source magnitude and filter to obtain adequate signal to noise ratio (SNR). Magnitude stacking technique is also used for multiple observations within a short span of a single night. At least two frames were observed every time in order to avoid the spurious photometric observations. 

The preprocessing involves the bad-pixel masking, bias subtraction, dark subtraction, flat correction, cosmic ray removal, astrometric solution which is done by the in-built BANZAI pipeline of LCO \citep{2018McCully}. We have downloaded the clean data from the LCO  server after all the corrections. The calibration, aperture, and PSF photometry were performed using \texttt{lcogtsnpipe} developed by \citet{2016Valenti}. The instrumental magnitudes in the B, V, r filters were derived from the difference images using point-spread function (PSF) photometry. To calibrate the data, nightly zero points and colour terms were established for each filter from the AAVSO Photometric All-Sky Survey (APASS) Data Release 9 catalogue \citep{2015Henden}. These calibration parameters were then applied to determine the calibrated GRB magnitude. To ensure accuracy, all detected stars in the field that matched with APASS were included, with sigma clipping performed to exclude potential variable stars. As GRB~250129A occurred very far away, it is unlikely to have host galaxy contribution at redshift z = 2.15. The LCO magnitudes in different filters are tabulated in Table \ref{tab:photometry}.

\subsection{ZEISS photometry of GRB~250129A}
\label{zeiss}

Optical observations of the field of GRB~250129A were conducted using the SAO RAS 1-meter telescope Zeiss-1000 equipped with one of the two instruments installed in the Cassegrain focus, CCD-photometer or focal reducer MAGIC. CCD-photometer is used to obtain direct images in the Johnson-Cousins B, V, Rc, Ic filters \citep{2020Komarov}. The nitrogen-cooled camera based on the EEV 42-40 CCD chip is used as a detector providing a $7^{\prime}.3 \times 7^{\prime}.3$ field of view with $2 \times 2$ binning mode and image scale of $0^{\prime\prime}.43/px$. The multimode focal reducer MAGIC with a liquid-cooled Andor iKon-L 936 CCD camera is used to obtain direct images in the Rc band providing $\sim 12^{\prime}$ circle of unvignetted field of view with $1\times1$ binning mode and image scale of $0^{\prime\prime}.45/px$ \citep{2022Afanasiev}.

The data were processed with the ESO-MIDAS standard procedures of bias subtraction, flat field correction, and cleaning of cosmic ray traces. To remove fringing patterns from Rc and Ic images, small shifts between individual frames were made. Sets of 300 seconds images in each band were used to create a median filtered image of the fringe pattern, which was subtracted from scientific images. Individual frames of each night were aligned into a common coordinate system and stacked for increasing the signal-to-noise ratio. The weather conditions during observations were good, typical seeing (FWHM of stellar-like objects) ranging from $1^{\prime\prime}.4$ to $2^{\prime\prime}.3$ allow performing aperture photometry without significant contamination from the nearby object located $\sim 5^{\prime\prime}$ North of the OT. 

The optical magnitudes are shown in Table \ref{tab:photometry} which were calibrated with the converted (\url{https://classic.sdss.org/dr4/algorithms/sdssUBVRITransform.php#Lupton2005}). Magnitudes of nearby reference stars from the Sloan Digital Sky Survey catalog (Table \ref{standards}). Initial magnitudes of the SDSS system were comparable with the APASS magnitudes of used reference stars (Fig. \ref{Fig:chart}).

\subsection{DOT and DFOT photometry of GRB~250129A}
\label{dot}

GRB~250129A was observed using the 2k $\times$ 2k Imager on the 1.3m Devasthal Fast Optical Telescope (DFOT) in U, B, V, R, I filters, and using the ADFOSC \citep{adfosc} instrument mounted on the 3.6m Devasthal Optical Telescope (DOT) in u, g, r, i, z filters. The Bias and Flat corrections were done following the standard procedures using \texttt{ccdproc} \citep{ccdproc}, and cosmic rays were removed using \texttt{astroscrappy} \citep{asctroscrappy}. The photometry was performed using a custom python photometry pipeline. The PSF was derived using \texttt{PSFEx} \citep{psfex_ascl} and the instrumental magnitudes were derived by fitting the detected sources with the PSF model using \texttt{photutils} \citep{photutils}. The final magnitudes were derived by calibrating the ugriz instrumental magnitudes against the SDSS DR17 \citep{sdss_dr17}, and the UBVRI instrumental magnitudes against the Gaia Synthetic Photometry Catalog \citep{gaia_synth_phot}. The calibrated magnitudes in each filter are presented in Table \ref{tab:photometry}. These calibrated magnitudes and their errors are converted to the flux density after correcting for the galactic extinction \citep{Schlafly2011}. 

\section{Spectroscopy and spectropolarimetry}
\label{ch4:specpol}

\subsection{DOT spectroscopy of GRB~250129A}
Low-resolution spectrum was taken using the ADFOSC mounted on the 3.6m DOT $\sim$ 18 hours since the burst. 5 consecutive exposures of 600s were obtained through a $1.5^{\prime\prime}$ slit and grism 676R (420 gr/mm). After image processing using \texttt{ccdproc} and astroscrappy, the individual exposures were stacked together. The spectrum was extracted and calibrated using standard routines in \texttt{IRAF} \citep{iraf}.  The wavelength calibration was done using the Hg-Ar lamp observed on the same night.

\begin{figure}
   \centering
   \includegraphics[width=\columnwidth]{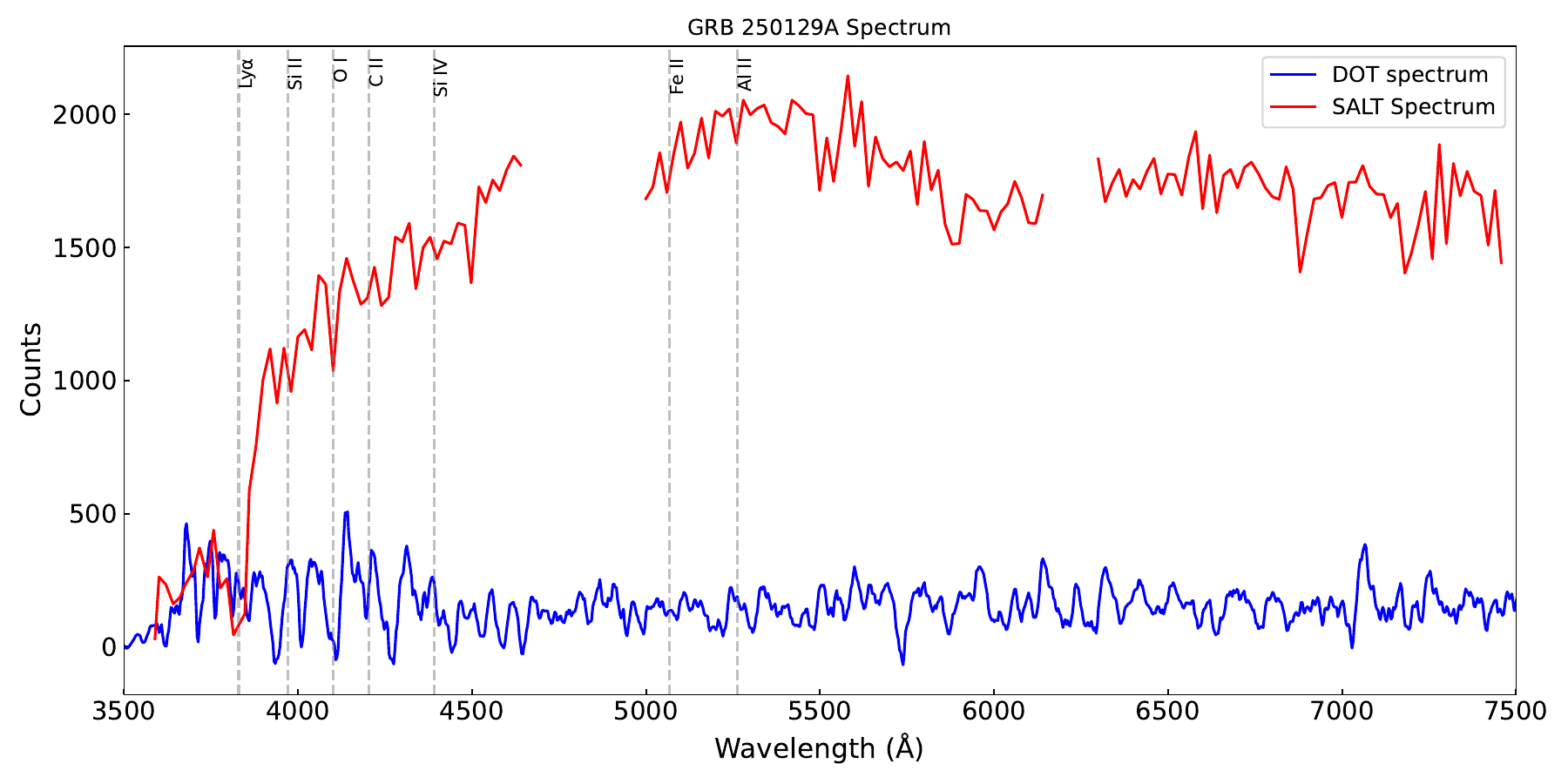}
      \caption{Optical spectrum of GRB~250129A using SALT and DOT which is converted to the rest-frame wavelength. The DOT spectrum is indicated with red whereas the SALT spectrum is colured in blue. Spectral lines at different wavelengths are also indicated. Note that the RSS spectra have been binned with 50\rm{\AA} bins.}
         \label{Fig:spec}
\end{figure}

\subsection{SALT spectroscopy of GRB~250129A}

The optical spectropolarimetry observation was taken with the Southern African Large Telescope \citep[SALT;][]{2006SPIE.6267E..0ZB} using the Robert Stobie Spectrograph \citep[RSS;][]{2003SPIE.4841.1463B, 2003SPIE.4841.1634K} situated at the South African Astronomical Observatory (SAAO). The RSS is capable of various modes of spectroscopy, including long-slit, multi-object slit, and Fabry-P\'erot.

The observations of GRB~250129A were taken as part of a community-wide transients monitoring campaign under the Proposal ID 2024-2-LSP-001. The RSS was set up in longslit spectropolarimetry \textsc{linear} mode with a $1.5^{\prime\prime}$ slit width, resulting in a resolution R of 560 to 900. The PG0700 grating was used at a 4.6$^{\circ}$ grating angle. Four observations (taken with the half-waveplate at angles 0$^{\circ}$, 45$^{\circ}$ , 22.5$^{\circ}$, and 67.5$^{\circ}$, respectively)  of 500\,s each were taken, and an arc frame was taken immediately after the science frames, using an Ar arc lamp. A field star was also observed at the same time (lying off centre on the slit) as an additional control on systematic errors.

The data was reduced using a modified version of the \textsc{pysalt/polsalt} data reduction pipeline  \citep{2010SPIE.7737E..25C, 2022heas.confE..56C}. The modification allowed for wavelength calibration to be performed using \textsc{iraf/noao},\footnote{\url{https://iraf.net/}}$^{,}$\footnote{\url{http://ast.noao.edu/data/software}} as well as enhanced cosmic ray cleaning using the python-based \textsc{lacosmic} package.

\section{Multi-wavelength modelling of GRB~250129A}
\label{ch5:model}

\subsection{FS and RS modelling}
\label{ch5:fsrs}

\begin{figure*}
    \centering
    \begin{minipage}[t]{0.49\textwidth}
        \centering
        \includegraphics[width=\linewidth]{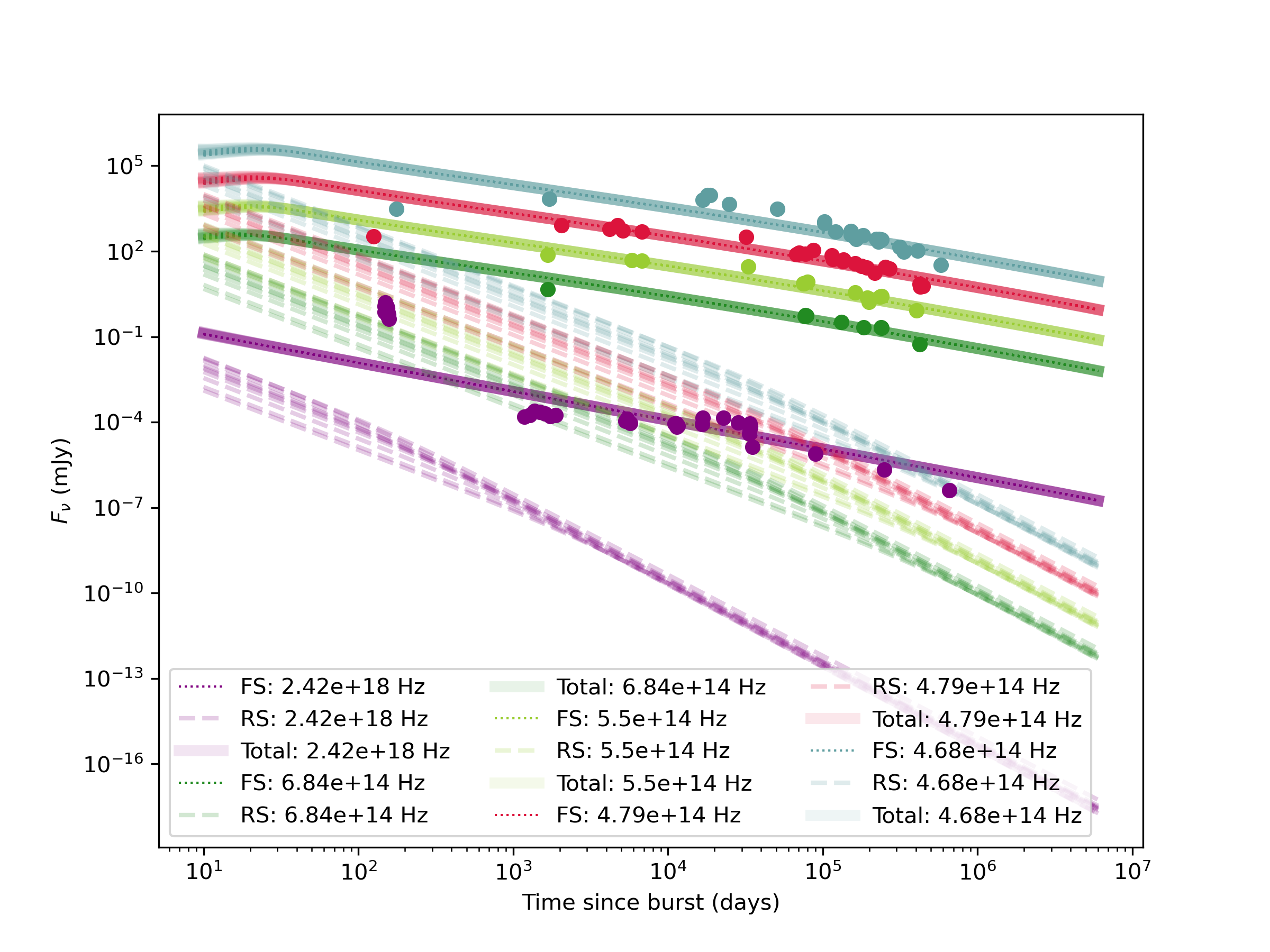}
        %\caption{ISM medium}
        %\label{fig:fig1}
    \end{minipage}
    \hfill
    \begin{minipage}[t]{0.49\textwidth}
        \centering
        \includegraphics[width=\linewidth]{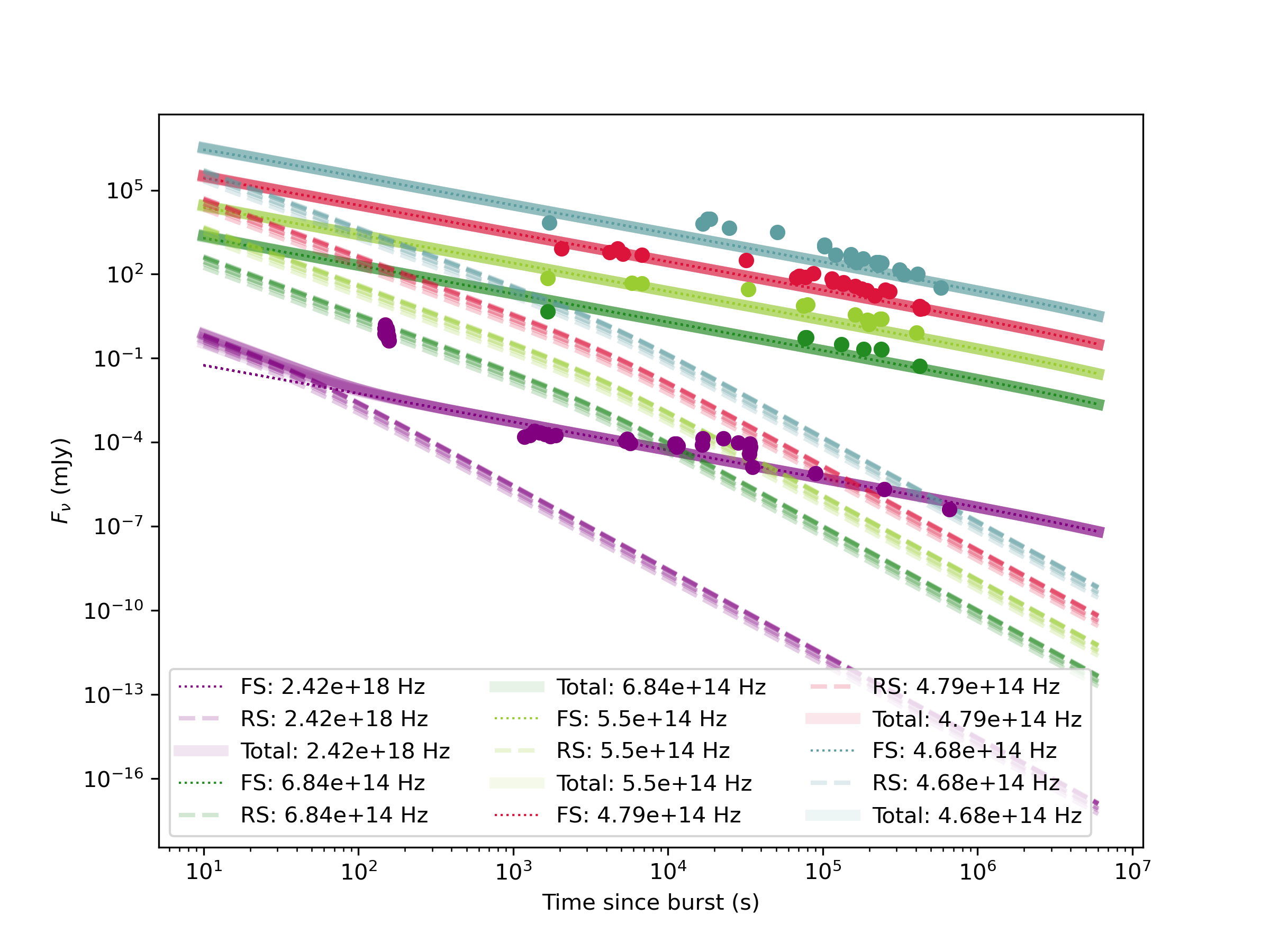}
        %\caption{Wind medium}
        %\label{fig:fig2}
    \end{minipage}
     \caption{Multi-wavelength light-curve modelling of GRB 250129A using a top-hat jet with combined FS+RS components. The top panel shows the best-fit model for a constant-density ISM environment, while the bottom panel illustrates the fit for a wind-like medium. The dataset includes \textit{Swift}-XRT X-ray observations and selected optical bands with the highest data coverage.}
    \label{fig:multi_panel}
\end{figure*}

The blast wave with initial Lorentz factor ($\Gamma_0$) and energy $E_0$ ploughs through the circumstellar medium having a density profile of $n(R)$ = $(A/m_p)R^{-k}$ and gets decelerated, which is responsible for the production of a two-shock system in GRB afterglow. The FS moves towards the unshocked ambient medium, whereas the RS propagates across the unshocked GRB ejecta. The RS significantly dominates the early afterglow depending on the magnetization parameter $\sigma$ = $B^2/4\pi n(R) m_p c^2$. In contrast, the FS dominates the afterglow emission after the time RS crosses the GRB ejecta. We considered two different ambient mediums, 1.\ constant density interstellar medium (ISM, k=0), 2.\ wind medium (k = 2).

\begin{table*} %[ht]
\centering
\caption{Best-fit parameters for GRB~250129A assuming a top-hat jet propagating through both ISM and wind-like media, modelled using the combined FS and RS framework.}
\begin{tabular}{lcccc}

%\toprule
\hline
Model & Prior range &  ISM medium & Wind medium\\
%\midrule
\hline\hline

$\theta_{\rm core}$ (rad)  & (0.035 - 0.261)   & $0.321^{+0.020}_{-0.020}$ & $0.256^{+0.001}_{-0.001}$ \\

$\log_{10} E_0$ (erg)   & (50 - 55)      & $54.446^{+0.023}_{-0.023}$ & $53.643^{+0.180}_{-0.140}$  \\
$\log_{10} n_{0}$(cm$^{-3}$) & ((-4) - 2) & $-0.110^{+0.180}_{-0.180}$ & -- \\
$\log_{10} A$(cm$^{-3}$) & ((-4) - 1)  & -- & $-0.185^{+0.005}_{-0.005}$  \\
$p$         & (1.9 - 2.6)                  & $2.012^{+0.001}_{-0.001}$ & $2.013^{+0.001}_{-0.001}$  \\
$\log_{10} \epsilon_e$  & ((-3) - (-1))      & $-1.360^{+0.023}_{-0.023}$ & $-2.070^{+0.016}_{-0.016}$\\
$\log_{10} \epsilon_B$   &  ((-5) - (-1))    & $-2.110^{+0.120}_{-0.120}$ & $-1.469^{+0.007}_{-0.007}$  \\
$p_{RS}$      & (1.9 - 2.6)                & $2.484^{+0.050}_{-0.050}$ & $2.001^{+0.001}_{-0.001}$  \\
$\log_{10} \epsilon_{e,RS}$ & ((-3) - (-1))         & $-2.884^{+0.087}_{-0.087}$ & $-2.652^{+0.019}_{-0.019}$\\
$\log_{10} \epsilon_{B,RS}$   & ((-6) - (-1))       & $-4.400^{+0.520}_{-0.520}$ & $-5.126^{+0.025}_{-0.025}$  \\

$\ln(Z)$        &  --              & $-3268.43 \pm 0.26$          & $-2138.89 \pm 0.45$        \\

\hline
\label{Tab:bestfit_fsrs}
%\bottomrule
\end{tabular}
\end{table*}

To model the multi-wavelength afterglow emission, we accounted for contributions from both the FS and RS, as well as their combined effects. Although the model also incorporates features such as jet break transitions and the onset of the Newtonian phase, these components did not significantly influence the results due to the absence of late-time observational data. The top-hat jet scenario with 9 model parameters were considered to explain the afterglow data. Nine parameters include the jet opening angle ($\theta_j$), isotropic equivalent energy ($E_{k, \rm iso}$), the ambient medium density ($n_0$), three FS parameters (p, $\epsilon_e$, $\epsilon_B$) and three RS parameters ($p_{RS}$, $\epsilon_{\rm e, RS}$, and $\epsilon_{\rm B,RS}$). The RS crossing time ($t_{\rm X}$) is considered in our model, which is fixed to 100 s for both the ISM and wind mediums.
%, respectively.

We attempted the Bayesian parameter estimation by the pyMultinest package based on the Nested Sampling Monte Carlo algorithm \textit{Mutinest} \citep{2009Feroz, 2014Buchner}. The nested sampling was performed for 5000 iterations. Prior ranges and their best fit values are tabulated in Table \ref{Tab:bestfit_fsrs}.

\subsection{Afterglow modelling with \textsc{Redback}/\textsc{Afterglowpy}}
\label{redback}

\begin{table*}
\centering
\caption{Best-fit parameters of GRB~250129A for different jet structure models using \textsc{Redback}/\textsc{Afterglowpy}. }
\begin{tabular}{lccc}

%\toprule
\hline
Model & Top hat & Gaussian core\\
%\midrule
\hline\hline
$\theta_{\rm obs}$ (deg) & $2.29^{+0.11}_{-0.12}$ & $2.86^{+0.11}_{-0.11}$  \\
$\theta_{\rm core}$ (deg)     & $1.72^{+0.11}_{-0.12}$ & $1.72^{+0.11}_{-0.12}$ \\
$\theta_{w}$ (deg)            & --                        & $1.57^{+0.01}_{-0.01}$ \\
$\log_{10} E_0$ (erg)         & $55.660^{+0.22}_{-0.460}$ & $53.900^{+0.180}_{-0.140}$  \\
$p$                           & $2.050^{+0.009}_{-0.009}$ & $2.190^{+0.010}_{-0.010}$  \\
$\log_{10} n_{\rm ism}$/cm$^{-3}$ & $1.250^{+0.220}_{-0.310}$ & $-0.470^{+0.210}_{-0.290}$  \\
$\log_{10} \epsilon_e$        & $-2.910^{+0.440}_{-0.210}$ & $-0.800^{+0.140}_{-0.180}$\\
$\log_{10} \epsilon_B$        & $-4.420^{+0.350}_{-0.220}$ & $-0.890^{+0.560}_{-0.410}$  \\
$\xi_N$                       & 1.0                        & 1.0                        \\
$\ln(Z)$                      & $-1808.37 \pm 0.08$          & $-1757.50 \pm 0.20$        \\

\hline
\label{Tab:bestfit_ap}
%\bottomrule
\end{tabular}
\end{table*}

We employed \textsc{Redback}/\textsc{Afterglowpy}\footnote{\url{https://github.com/geoffryan/afterglowpy}} to model the multi-wavelength afterglow emission. Our primary aim was to constrain the jet opening angle, viewing angle, and the wing contribution to the afterglow. We considered different jet structures (Top hat, Gaussian, and Gaussian core) available in the code. The fitting of GRB~250129A was performed using the \textsc{PyMultinest} sampler embedded within \textsc{Redback} by \citet{2024MNRAS.531.1203S} and coupled with \textsc{Afterglowpy} models by \citet{2020ApJ...896..166R}. In the Gaussian core scenario, the emission is governed by eight key parameters: - $\theta_{\rm obs}$ : viewing angle of the observer, $E_0$: isotropic equivalent energy, $\theta_{\rm core}$: half opening angle of the jet, $\theta_{\rm wing}$: outer truncation angle, $n_0$: number density of the ambient medium, $p$: electron power-law index and fraction of energy distributed to accelerating electrons and magnetic field $\epsilon_e$ and $\epsilon_B$. For the top hat jet configuration, contribution from $\theta_{\rm wing}$ was excluded. The electron acceleration fraction ($\xi_e$) was fixed at 1.0, and a redshift of $z = 2.151$ was adopted. We kept the initial Lorentz factor $\Gamma_0$ fixed at 500, calculated from the formulations of \citet{2021MNRAS.505.1718J, 2025arXiv250518041B}. The priors employed in the analysis were uniform in energy, core angle, and circumburst density, while a sine function prior was adopted for the viewing angle. We explored all the three models in \textsc{Afterglowpy} to model GRB~250129A.

\subsection{Flaring}
\label{flare}
Beyond the standard afterglow emission, we identify a clear excess emission which is evident in the X-ray band, that is an indication of prolonged central engine activity. This flaring activity in X-ray band, persisting from $10^4$ to $4 \times 10^4$ seconds since the explosion, can be well described with the phenomenological model like Norris function.  
This function have asymmetric profiles including separate rise and decline indices ($\tau_1$ and $\tau_2$, respectively). The intensity of the flare can be formulated by \citet{2005ApJ...627..324N},
\begin{equation}
    I(t) = A \lambda\ \bigg(-\frac{\tau_1}{(t-t_i)} - \frac{(t-t_i)}{\tau_2}\bigg)\,,
\end{equation}
where $t_{\rm i}$ is the onset time of the flaring emission, and the equation holds for t > $t_{\rm i}$. We then adjust flux density values corresponding to the flare intensity. Comparing the flaring observational data with the Norris model, the $t_i = 1.5 \times 10^3$ s. Using the formulations given by \cite{2005ApJ...627..324N}, the peak time $t_{\rm peak} = 1.80 \times 10^{4}$ s which is in close agreement with the observed flare maximum.

\section{Results and analysis}
\label{ch6:res}

\subsection{Spectropolarimetric results}

   \begin{figure*}
   \centering
   \includegraphics[width=\textwidth]{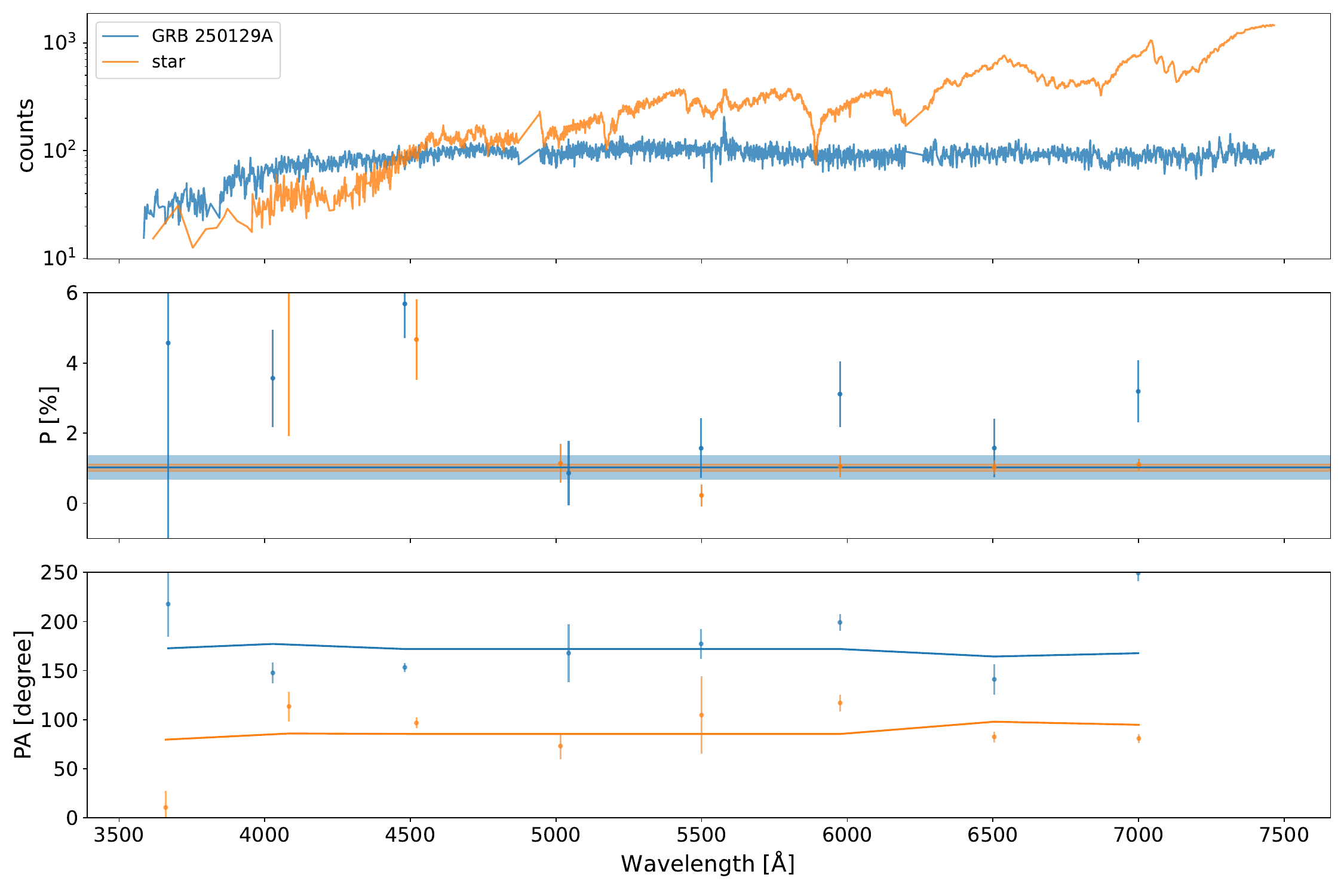}
      \caption{The polarization of GRB 250129A (blue) compared to a comparison star (orange) observed by SALT $\sim$ 19 hours since the burst. The plot shows the unbinned counts (top panel), and the degree of polarization (middle panel), and the position angle (bottom panel) binned on 500\,\AA{}. The middle panel shows the error in the mean value, and the standard deviation of the polarization measurements between 4500 and 7000\,\AA{} as shaded regions (see text for details). In the bottom panel the solid lines show the box car average of the polarization averaged over 10 data points.}
         \label{Fig:specpol}
   \end{figure*}

The spectropolarimetric data obtained $\sim$ 19 hours post-burst shows hint of late-time polarisation, with the PD of $\sim$ 1.03\%. There is no change in polarisation angle (PA) has been noticed across the observed frequency range as shown in the second and third panels of Fig.~\ref{Fig:specpol}. The probable origin of this late-time polarisation observation in GRB~250129A is investigated through multiple afterglow models. The signal to noise ratio for the spectropolarimetric observation is poor, so that we had to choose larger wavelength binning (500 \AA) to decrease the errors.

We accounted for three sources of foreground absorption that contribute to the observed polarisation: the milky way, the GRB host galaxy, and intervening absorbers along the line of sight. The host-galaxy polarization contribution was constrained using the extinction inferred from the afterglow spectral energy distribution, adopting the typical value of $A_V^{\rm host} \simeq 0.2$\,mag, which implies an upper limit of $P_{\rm host}^{\max} \lesssim 0.6\%$. At $z = 2.151$, the observed optical band corresponds to rest-frame ultraviolet wavelengths ($\lambda_{\rm rest} \sim 6000\,\AA$), where dust-induced polarisation is inefficient, constraining the realistic host contribution to $\lesssim 0.3\%$. The contribution to polarisation from galactic extinction can be estimated using $A_V = 0.017$ and $E(B-V) = 0.0387$ for milky way, which is estimated to be $9 \times E(B-V) = 0.33\%$. Considering MgII and Damped Lyman-alpha (DLA) systems ($E(B-V) \leq 0.05$), the maximum polarisation contribution from the intervening absorbers along the line of sight is 0.45\%. For random magnetic-field orientations and vector cancellation in Stokes space, the net polarization from intervening absorbers is expected  to be 0.2\%. The total foreground polarization contribution is 0.8\%, lower than the observed PD of GRB~250129A, which provides a hint of intrinsic origin of polarization for GRB~250129A.

\begin{figure}   
\begin{center}
\includegraphics[width=\columnwidth]{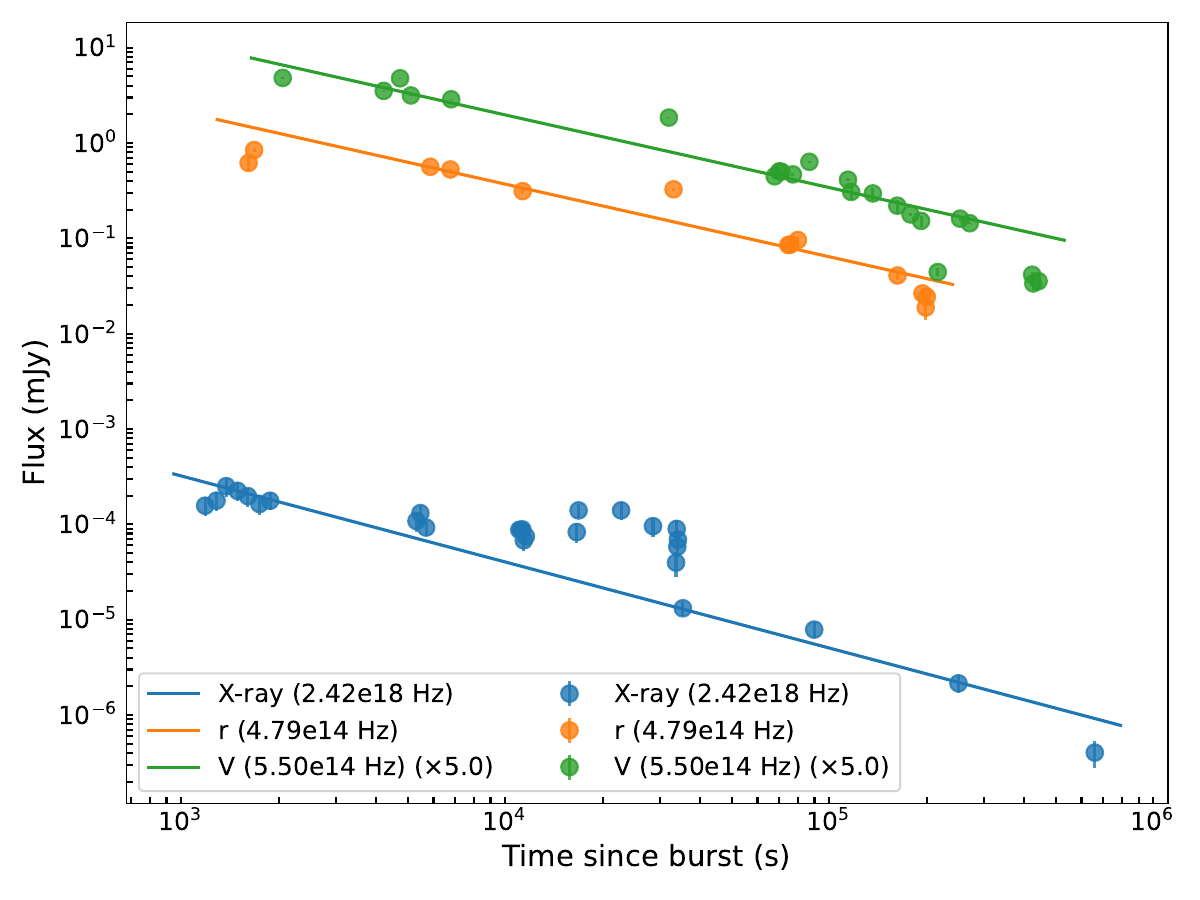}
\caption{
%with $\chi^2$ optimisation. 
Light curves of GRB 250129A fitted with phenomenological single power-law model. Blue circles show the X-ray flux at 10 keV, while green and orange circles correspond to the optical V and
r bands, respectively. Solid lines denote the best-fit models, with an index of $\alpha_X = 0.903 \pm 0.021$ for
X-ray and $\alpha_o = 0.792 \pm 0.002$ for the optical bands. \label{fig:plfit}}
\end{center}   
\end{figure}
\subsection{Multi wavelength light curve analysis}

To examine whether the model is fitting well with the multi-wavelength dataset, we first fit the X-ray and optical light curves with the standard phenomenological power-law model. Fig.~\ref{fig:plfit} shows slightly different decay indices of $\alpha_{\rm o} = 0.762 \pm 0.002$ and $\alpha_{\rm X} = 0.903 \pm 0.021$, respectively, for the optical and X-ray bands. Although, the difference is insufficient to robustly establish the presence of a cooling break between the optical and X-ray bands, the flux-decay index ($\alpha \sim 0.7 -0.9$) from the fitting indicate a typical slow-cooling spectrum. We compare the flux-decay indices with the afterglow closure relations for a FS-dominated synchrotron model \cite{2018ApJ...866..162G}. The results clearly prefer a constant density interstellar medium (ISM) over a wind-like environment, with the electron's spectral index p $\sim$ 2.1, and the observed optical and X-ray frequencies ($\nu_{\rm obs}$) satisfying the relation $\nu_{\rm m} < \nu_{\rm obs} < \nu_{\rm c}$. Here $\nu_{\rm m}$ and $\nu_{\rm c}$ are frequencies of synchrotron radiation by electrons with the minimum energy and cooling energy, respectively. 

To interpret the panchromatic dataset and the hint of PD in GRB~250129A, we explored two distinct modelling approaches: 1. the standard blast wave model with the contributions from FS and RS emission, 2. Standard afterglow model considering the jet geometric effects associated with the observer’s viewing angle and flaring activity. The first model does not incorporate the jet structure or the observer’s viewing geometry. While the second model, accounting for geometric effects, does not include a contribution from the RS component. Firstly, we employed our custom-developed afterglow emission model in which a spherical jet is propagating through uniform density ISM or wind-like environments, incorporating both the FS and RS components. The details of the modelling are outlined in section \ref{fsrs}. The RS crossing time ($t_{\rm X}$) was calculated to determine the dominant emission component at each epoch. It is clearly evident that the broadband afterglow emission of GRB~250129A is dominated by FS from early to late times, with RS contributing marginally only at very early epochs ($\lesssim$100 s) in X-ray. The wind-like medium provides a significantly better fit to the data compared to the uniform ISM case based on the log(Z) value in Table \ref{Tab:bestfit_fsrs}. The modelling yields high isotropic-equivalent kinetic energy of $4.4 \times 10^{53}$ erg and suggests a relatively low circumburst density of $\sim 0.95\,\mathrm{cm^{-3}}$. 

\begin{figure}
   \centering
   \includegraphics[width=\linewidth]{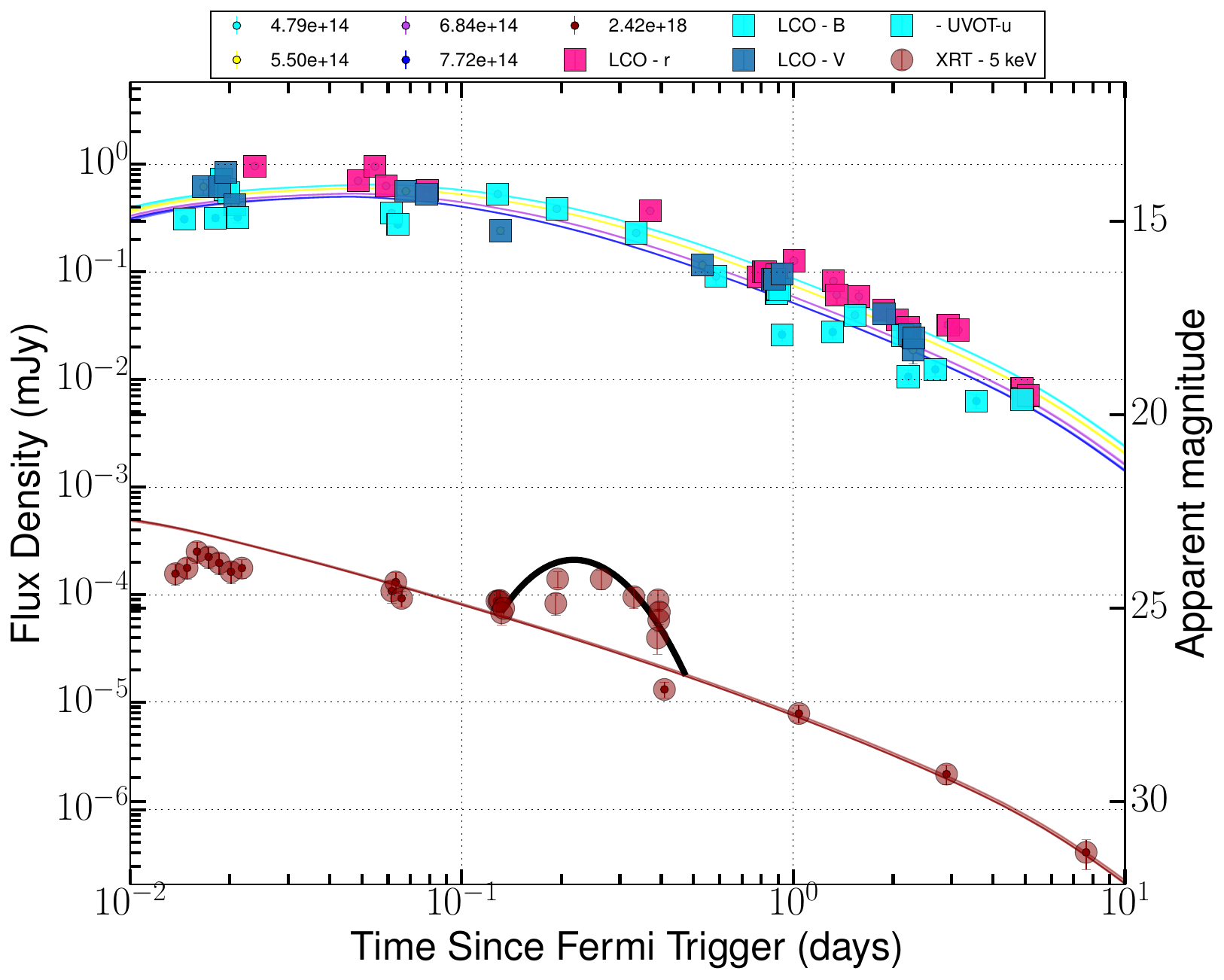}
      \caption{Broadband afterglow modelling of GRB2~250129A using \textsc{Redback}/\textsc{Afterglowpy} for the jet with Gaussian core and wing component in the homogeneous ISM environment. The best fit values for X-ray and optical bands are denoted with solid lines. Black solid line indicates the flaring emission in X-ray band.}
         \label{Fig:fit_ap}
\end{figure}

We again explored three jet scenarios (Top-hat jet, Gaussian jet, and Gaussian-core) of \textsc{Redback}/\textsc{Afterglowpy} to model the afterglow data of GRB~250129A. The Gaussian jet failed to reproduce the observed broadband data, whereas both the Gaussian-core with wing component and top-hat jet models propagating in a homogeneous ISM provided satisfactory fits. Based on the lower log-evidence (ln Z) value mentioned in Table \ref{Tab:bestfit_ap}, we identify the jet with Gaussian core with wing component as the best-fit model. The forest plot and posterior distributions for Gaussian core jet with wing contribution are shown in Figs.~\ref{Fig:fit_ap} and \ref{Fig:coner_ap}. The core and viewing angle of this scenario suggest a narrow jet with $\theta_{\rm core} = 1.71^{\circ}\pm 0.11$, viewing angle $\theta_{\rm obs} = 2.86^{\circ}\pm 0.11$, and the wing angle $\theta_{\rm wing} = 89.95^{\circ}\pm 6.25$. GRB~250129A is a highly energetic GRB with inferred isotropic energy of $7.9 \times 10^{53}$ erg and low ambient medium density ($n_0$ = 0.33 cm$^{-3}$). The energetic nature of the afterglow can be explained through the narrow jet. Afterglow modelling suggests that $\theta_{\rm obs}$ > $\theta_{\rm core}$, which clearly indicates an off-axis emission with significant contribution coming from the wing component.

\subsection{Polarimetric modelling}

\begin{figure}   
\begin{center}
\includegraphics[width=\columnwidth]{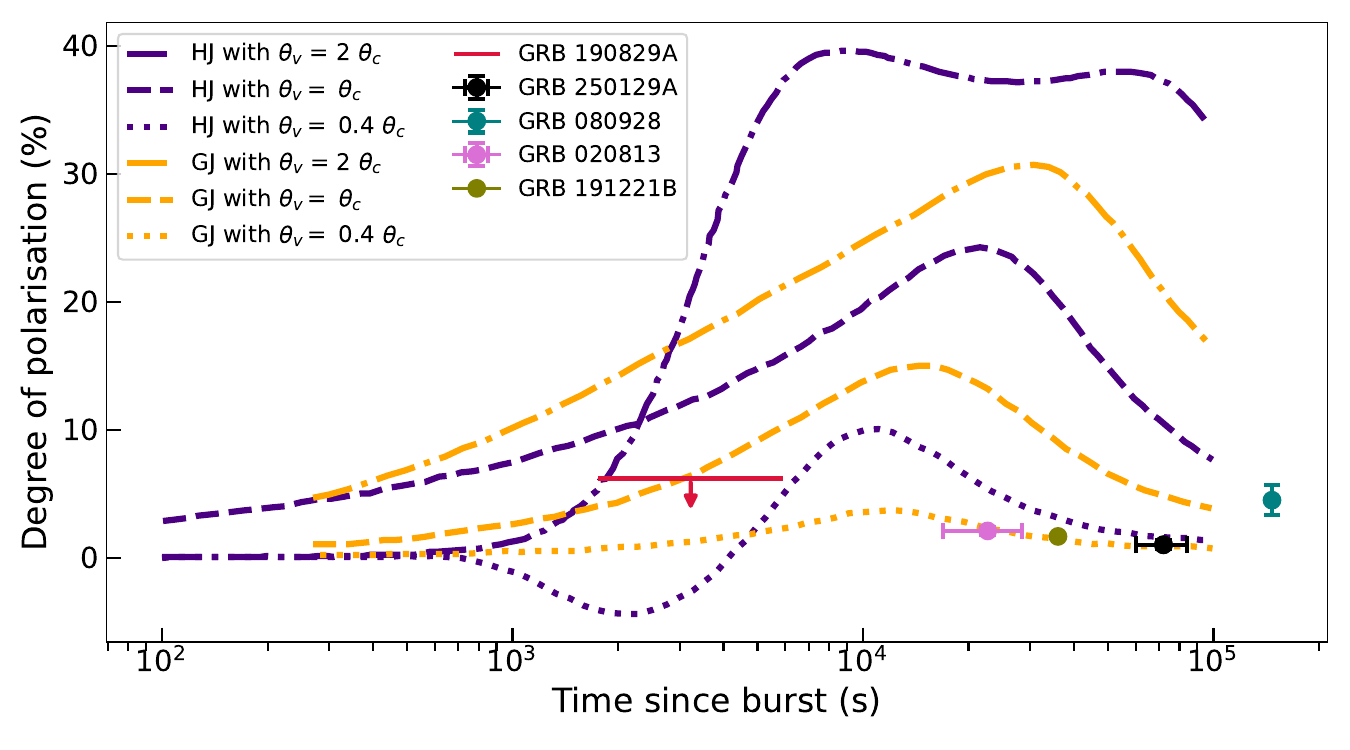}
\caption{
The polarisation model adopted from \protect\cite{2004MNRAS.354...86R,2022MNRAS.512.2337D} for different jet structures and geometry. The black dot represents the late-time polarisation measurement of GRB~250129A obtained at $\sim$ 19 hours after the burst. The degree of linear polarisation of GRB~250129A is compared with previously observed GRB sample (GRB~020813 \protect\cite{2003ApJ...584L..47B}, GRB~080928 \protect\cite{2022A&A...666A.179B}, GRB~190829A \protect\cite{2022MNRAS.512.2337D}, GRB~191221B \protect\cite{2021MNRAS.506.4621B}) with the polarisation detection through spectropolarimetry. %\textcolor{magenta}{SR: Can you extended the model curves to the time covering the observation of GRB~080928?}
}
\label{fig:polmod}
\end{center}   
\end{figure}

The degree of linear polarisation with time for GRB~250129A was compared with four other GRBs (GRB~020813 \citep{2003ApJ...584L..47B}, GRB~030329A \citep{2003Natur.426..157G}, GRB~191221B \citep{2021MNRAS.506.4621B}, GRB~080928 \citep{2022A&A...666A.179B}.) observed through spectropolarimeter, as illustrated in Fig.~\ref{fig:polmod}. It also shows the time-dependent evolution of the PD for GRBs, for both top-hat and Gaussian jet structures. The polarisation curves were generated following the prescription given by  \cite{2004MNRAS.354...86R,2022MNRAS.512.2337D}, which provide a framework to explore how jet structure and viewing configuration shape the observed signal. Three distinct cases 1. on-axis ($\theta_v$ = 0.4$\times \theta_c$), 2. transition ($\theta_v$ = $\theta_c$), and 3. off-axis ($\theta_v$ = 2$\times \theta_c$) were considered. GRB~250129A is following the Gaussian jet with moderate lateral expansion and the observing angle is lying within the jet opening angle. Although it does not show a clear distinction between the Gaussian and top-hat jet. 
It is also evident that the configuration for the polarisation model has discrepancy with the afterglow model that indicates $\theta_{\rm obs} > \theta_{\rm core}$. 
Therefore, no concrete statement can be made about the jet structure from the polarisation observation. 

\section{Summary}
\label{ch7:conclu}
We present a comprehensive multi-epoch broadband photometric and late-time spectropolarimetric study of GRB~250129A. GRB~250129A, occurred at high redshift (z = 2.151), is a high energetic GRB, exhibiting a very bright early optical emission. We initiated a panchromatic follow-up observation campaign using several telescopes globally, such as LCO global telescope network, Swift-UVOT, SAO-RAS ZEISS-1000 telescope, 3.6 m DOT, and 1.3 m DFOT. The earliest detection by the 0.4 m LCO telescope reveals a bright optical afterglow (r = 17 mag) at $\sim$ 1 hour since the explosion. Our dataset is supplemented with the multi band optical data reported in the GCN circulars and the X-ray light curve was retrieved from the Swift XRT repository \footnote{\url{https://www.swift.ac.uk/xrt_curves/01285812/}}. Spectropolarimetric and spectroscopic observations were conducted using the SALT/RSS and DOT/ADFOSC instruments, respectively. We obtained the spectropolarimetric observations $\sim$ 19 hours since the explosion. The hint of polarisation enabled us to investigate its physical origin using the high-cadence, multi-wavelength photometric coverage. This synergy between photometry and spectropolarimetry provides unique insight into the burst and environment properties og GRB~250129A.

Detection of polarisation in GRB afterglows is very rare, with measurements reported for only a few dozen events. In most of GRBs, high polarisation can be observed during the early afterglow phase due to RS, while FS regime at late times is typically expected to exhibit only weak polarisation. Polarisation detection through spectropolarimetry even uncommon, yet provides additional information about the break frequencies and the dust polarisation due to milky way or host. To date, only five GRBs (GRB~020813 \citep{2003ApJ...584L..47B}, GRB~021004 \citep{2003ApJ...592..457W}, GRB~030329A \citep{2003Natur.426..157G}, GRB~191221B \citep{2021MNRAS.506.4621B}, GRB~080928 \citep{2022A&A...666A.179B}.) have yielded spectropolarimetric measurements during their afterglows, with low PD. High PD was quoted for GRB~030329A at late-phase during the supernova. The polarisation observation of GRB~250129A shows the hint of polarisation at very late-time. But unfortunately there is no change in PA with wavelength.

We explore a standard afterglow models to explain the broadband afterglow and hint of polarisation using multi-wavelength dataset. Standard blastwave model with a combination of FS and RS was firstly considered to check the presence of RS component at late-time. As the RS dominates during very early afterglow phase of GRB~250129A, it is challenging to explain the PD at late-time with RS. Due to unavailability of very early time spectropolarimetric observation, we discard the presence of RS. Alternate scenarios, such as the flaring due to prolonged central engine is also explored which evident in X-ray and r band light curves. But the flaring internal and spectropolarimetric observations do not coincide with each other. Spectropolarimetric observation during the flaring time might make a solid conclusion that the PD is coming from the flaring. Now we considered the off-axis scenario where the $\theta_{\rm obs} > \theta_{\rm core}$. Multi-wavelength modelling with the Gaussian jet structure and additional wing component of photometric data clearly indicates GRB~250129A is a high energetic off-axis GRB occurred in a low-dense ambient medium. In case of off-axis GRBs, the symmetry of the magnetic field domains break, and we expect the hint of at late-time. Another puzzling property inferred from the afterglow modelling is $\epsilon_B = 0.12$ which is higher than the standard GRBs, possibly indicating a magnetically dominated outflow. The afterglow modelling of GRBs constrains the $\epsilon_B$ between $10^{-6} - 10^{-1}$ \citep{2014Santana, 2016Laskar, Aksulu2022}. At late times, when the outflow spreads and the $\Gamma$ decreases, a higher $\epsilon_B$ enables even the afterglow emission remains strong enough to carry the polarisation signature, making it detectable over a longer period.

\section*{Acknowledgements}

This research was supported by the National Research Foundation (NRF) of South Africa through a BRICS Multilateral Grant with number 150504 to SR. AG was supported by NRF  through a Postdoctoral Fellowship. RG was sponsored by the National Aeronautics and Space Administration (NASA) through a contract with ORAU. The views and conclusions contained in this document are those of the authors and should not be interpreted as representing the official policies, either expressed or implied, of the National Aeronautics and Space Administration (NASA) or the U.S. Government. The U.S. Government is authorized to reproduce and distribute reprints for Government purposes notwithstanding any copyright notation herein. I would like to sincerely thank the team of SALT, DOT, and the Zeiss-1000 telescope for performing the prompt observation of GRB~250129A afterglow. 

%%%%%%%%%%%%%%%%%%%%%%%%%%%%%%%%%%%%%%%%%%%%%%%%%%
\section*{Data Availability}

All data used in this work are publicly available from other papers and the \textit{Swift} website \url{https://www.swift.ac.uk/}. The optical data underlying this article is available in the article.

%%%%%%%%%%%%%%%%%%%% REFERENCES %%%%%%%%%%%%%%%%%%

% The best way to enter references is to use BibTeX:

\bibliographystyle{mnras}
\bibliography{example} % if your bibtex file is called example.bib

% Alternatively you could enter them by hand, like this:
% This method is tedious and prone to error if you have lots of references
%\begin{thebibliography}{99}
%\bibitem[\protect\citeauthoryear{Author}{2012}]{Author2012}
%Author A.~N., 2013, Journal of Improbable Astronomy, 1, 1
%\bibitem[\protect\citeauthoryear{Others}{2013}]{Others2013}
%Others S., 2012, Journal of Interesting Stuff, 17, 198
%\end{thebibliography}

%%%%%%%%%%%%%%%%%%%%%%%%%%%%%%%%%%%%%%%%%%%%%%%%%%

%%%%%%%%%%%%%%%%% APPENDICES %%%%%%%%%%%%%%%%%%%%%

\appendix

\section{Some extra material}

\subsection{Theoretical framework of FS - RS modelling}
\label{fsrs}
Synchrotron emission is characterised by three break frequencies: the frequency with minimum Lorentz factor of accelerated electron $\nu_m$, cooling frequency $\nu_c$, and self-absorption frequency $\nu_a$. The time-dependence of $\nu_m$ and $\nu_c$ can be estimated from the evolution of the bulk Lorentz factor ($\Gamma$). 
%For a constant density ISM $B' \propto \Gamma \propto t_{\text{\rm obs}}^{-3/8}, \quad \gamma_m \propto \Gamma \propto t_{\text{\rm obs}}^{-3/8}, \quad \text{and so } \nu_m \propto B' \gamma_m^2 \Gamma \propto \Gamma^4 \propto t_{\text{\rm obs}}^{-3/2}$ and similarly for a wind medium $\nu_c \propto t_{\text{\rm obs}}^{-3/2}$. 
The expressions of $\nu_m$ and $\nu_c$ are adopted from \citet{2002Granot, 2013bGao}. For the constant density ISM,
\begin{equation}
\begin{aligned}
\nu_m &= 3.3 \times 10^{14}~\mathrm{Hz}\,(1+z)^{1/2}\,
\epsilon_{B,-2}^{1/2}\left[\epsilon_e g(p)\right]^2
E_{52}^{1/2} t_{\rm obs,d}^{-3/2}, \\
\nu_c &= 6.3 \times 10^{15}~\mathrm{Hz}\,(1+z)^{-1/2}\,
\epsilon_{B,-2}^{-3/2} E_{52}^{-1/2} n_p^{-1}
t_{\rm obs,d}^{-1/2},
\end{aligned}
\end{equation}
where, $\epsilon_e$ and $\epsilon_B$ are the equipartition of energy going to the accelerating electron and the magnetic field respectively, the dimensionless quantity $g(p) = p-2/p-1$ for p $>$ 2, and $t_{obs,d}$ is the time in the observer frame in day unit. For a wind-like medium with $k=2$, the expression changes vastly as,
\begin{align}
\nu_m &= 4.0 \times 10^{14} \, \text{Hz} \,
(p - 0.69)(1 + z)^{1/2} \epsilon_{B,-2}^{1/2} \left[ \epsilon_e g(p)  \right]^2 E_{52}^{1/2}  
 t_{\text{obs},d}^{-3/2}, \ and \\[6pt]
\nu_c &= 4.4 \times 10^{13} \, \text{Hz} \,
(3.45 - p) \exp(0.45p)(1 + z)^{-3/2} \epsilon_{B,-2}^{-3/2} E_{52}^{1/2} 
A_*^{-2} t_{\text{obs},d}^{1/2},
\end{align}
where the density of the wind medium $n(R) = A_*(3 \times 10^{35}) R^{-2} \quad cm^{-3}$. $A_*$ becomes unity when the mass loss rate and wind speed become $10^{-5} M_{\odot}~{\rm yr}^{-1}$ and $10^{8}$ cm/s, respectively. The self-absorption frequency $\nu_a$ is a time-independent quantity, which can be derived from \citet{2000Panaitescu} as,
\begin{equation}
\begin{aligned}
\nu_a &= 1.0 \times 10^{8}~\mathrm{Hz}~
\epsilon_{B,-2}^{1/5}\,\epsilon_e^{-1}\,E_{52}^{1/5}\,n_0^{3/5},
&\quad \text{(ISM)} \\[6pt]
\nu_a &= 1.0 \times 10^{9}~\mathrm{Hz}~
\epsilon_{B,-2}^{1/5}\,\epsilon_e^{-1}\,E_{52}^{-2/5}\,A_*^{6/5}\,t_{{\rm obs},d}^{-3/5}.
&\quad \text{(Wind)}
\end{aligned}
\end{equation}
The peak flux of the synchrotron spectrum can be written as, 
%\src{Please check the following equations with eq.18, eq.B22 and eq.B45 in \cite{2021MNRAS.505.1718J}. I see some differences.}
%
\begin{equation}
    f_{\nu,max} = (1+z)\frac{N_{tot}P_{\nu,max}^{\prime} \Gamma}{4\pi D_L^2}\,,
\end{equation}

where $N_{\rm tot}$ is the total number of electrons contributing to resulting radiation at frequency $\nu$ and  $P_{\nu,max}$ is the power radiated per unit frequency for one electron $P_{\nu,max} = \sqrt{3}q^3B/m_e c^2$. For ISM $F_{\nu, \rm max} \propto N_{tot}B\Gamma \propto R^3\Gamma^2$ which is time-independent, and for a wind-like medium $F_{\nu, \rm max} \ \propto R\Gamma \ \propto \Gamma^2 \propto t_{\rm obs}^{-1/2}$. Following \citet{2002Granot} and \citet{2003Yost}, peak specific flux can be calculated from ISM and Wind mediums, respectively, as
\begin{align}
f_{\nu,\max} &= 1.6~\text{mJy}~(1+z)\,\epsilon_{B,-2}^{1/2} 
E_{52} n_p^{1/2} D_{L,28}^{-2}, 
\ \text{for ISM} \nonumber \\
f_{\nu,\max} &= 7.7~\text{mJy}~(p+0.12)(1+z)^{3/2} 
\epsilon_{B,-2}^{1/2} E_{52} A_* D_{L,28}^{-2} 
t_{\text{obs,d}}^{-1/2},\ \text{for Wind}.
\end{align}

For the RS emission, we calculated the break frequencies and the peak flux using the formulations given by \citet{2013bGao}.

\begin{equation}
\begin{aligned}[t]
\nu_m &= 8.5 \times 10^{11}~\text{Hz}~ z^{19/35}
\left( \frac{g(p)}{g(2.3)} \right)
E_{52}^{18/35} \Gamma_{0,2}^{-74/35}
n_{0,0}^{-1/70} \epsilon_{e,-1}^{2}
\epsilon_{B,-2}^{1/2} t_2^{-54/35}, \\[4pt]
\nu_{\rm cut} &= 4.3 \times 10^{16}~\text{Hz}~ z^{19/35}
E_{52}^{-16/105} \Gamma_{0,2}^{-292/105}
n_{0,0}^{-283/210} \epsilon_{B,-2}^{-3/2}
t_2^{-54/35}, \\[4pt]
F_{\nu,\max} &= 7.0 \times 10^5~\mu\text{Jy}~ z^{69/35}
E_{52}^{139/105} \Gamma_{0,2}^{-167/105}
n_{0,0}^{37/210} \epsilon_{B,-2}^{1/2}
D_{28}^{-2} t_2^{-34/35}, \\[4pt]
\nu_a &= 1.4 \times 10^{13}~\text{Hz}~ z^{-73/175}
\left( \frac{g(p)}{g(2.3)} \right)
E_{52}^{69/175} \Gamma_{0,2}^{8/175}
n_{0,0}^{71/175} \epsilon_{e,-1}^{-1}
\epsilon_{B,-2}^{1/5} t_2^{-102/175}, \\[-2pt]
&\hspace{-1.5em}(\nu_a < \nu_m < \nu_c) \\[4pt]
\nu_a &= 3.7 \times 10^{12}~\text{Hz}~
z^{\frac{19p - 36}{35(p+4)}}
\left( \frac{g(p)}{g(2.3)} \right)
E_{52}^{\frac{2(p+29)}{35(p+4)}}
\Gamma_{0,2}^{\frac{-74p - 44}{35(p+4)}}
n_{0,0}^{\frac{94 - 70p}{70(p+4)}}
\epsilon_{e,-1}^{\frac{2(p - 1)}{p+4}}
\epsilon_{B,-2}^{\frac{p + 2}{2(p+4)}}\\
%t_2^{\frac{-54p + 104}{35(p+4)}}, \\[-2pt]
&\hspace{-1.5em}t_2^{\frac{-54p + 104}{35(p+4)}}, (\nu_m < \nu_a < \nu_c)
\end{aligned}
\end{equation}
where $\nu_{\rm cut}$ indicates the cut-off frequency of the synchrotron spectrum  which is distinct from the synchrotron cooling frequency $\nu_{\rm c}$. No electrons were accelerated after the RS crossing time and $\nu_{\rm cut}$ was calculated from $\nu_{\rm c}$ at $t_{\rm X}$ due to adiabatic expansion \citep{2000ApJ...545..807K}.

\begin{figure*}
   \centering
   \includegraphics[width=0.8\textwidth]{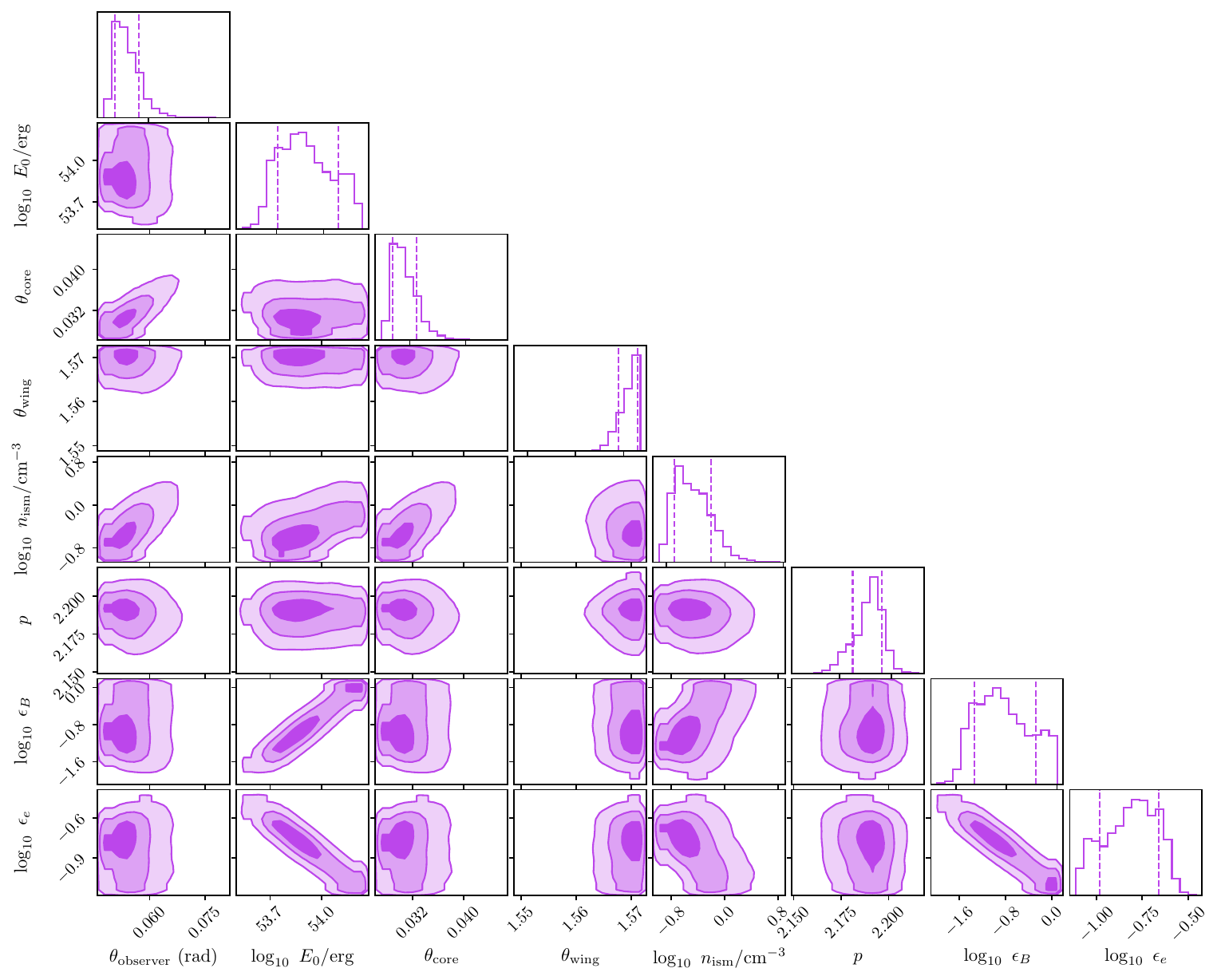}
      \caption{Corner plots for ISM fit using \textsc{Redback}.}
         \label{Fig:coner_ap}
\end{figure*}

\begin{figure*}
   \centering
   \includegraphics[width=0.9\textwidth]{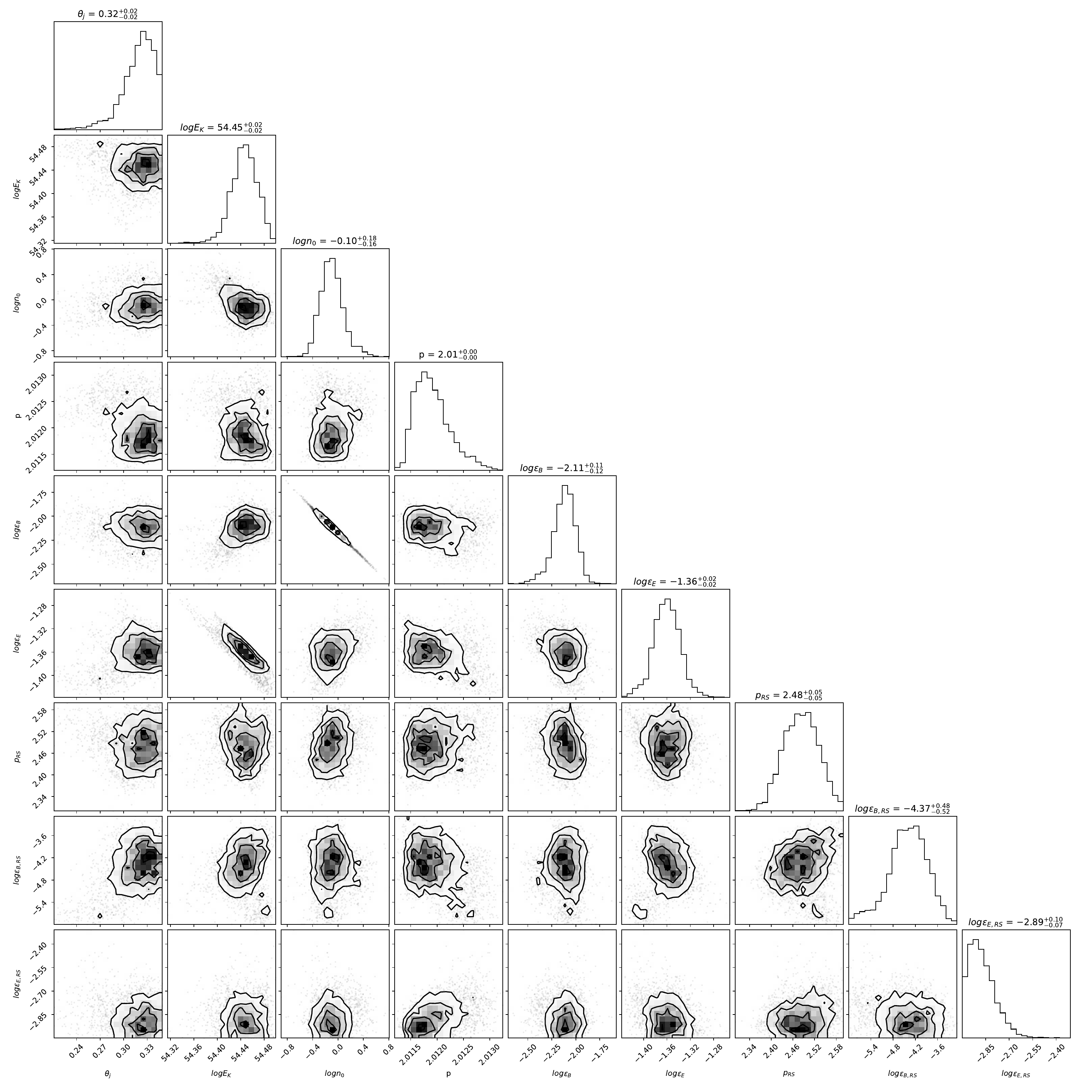}
      \caption{Corner plots for ISM fit.}
         \label{Fig:ISM_fsrs}
\end{figure*}

\begin{figure*}
   \centering
   \includegraphics[width=0.9\textwidth]{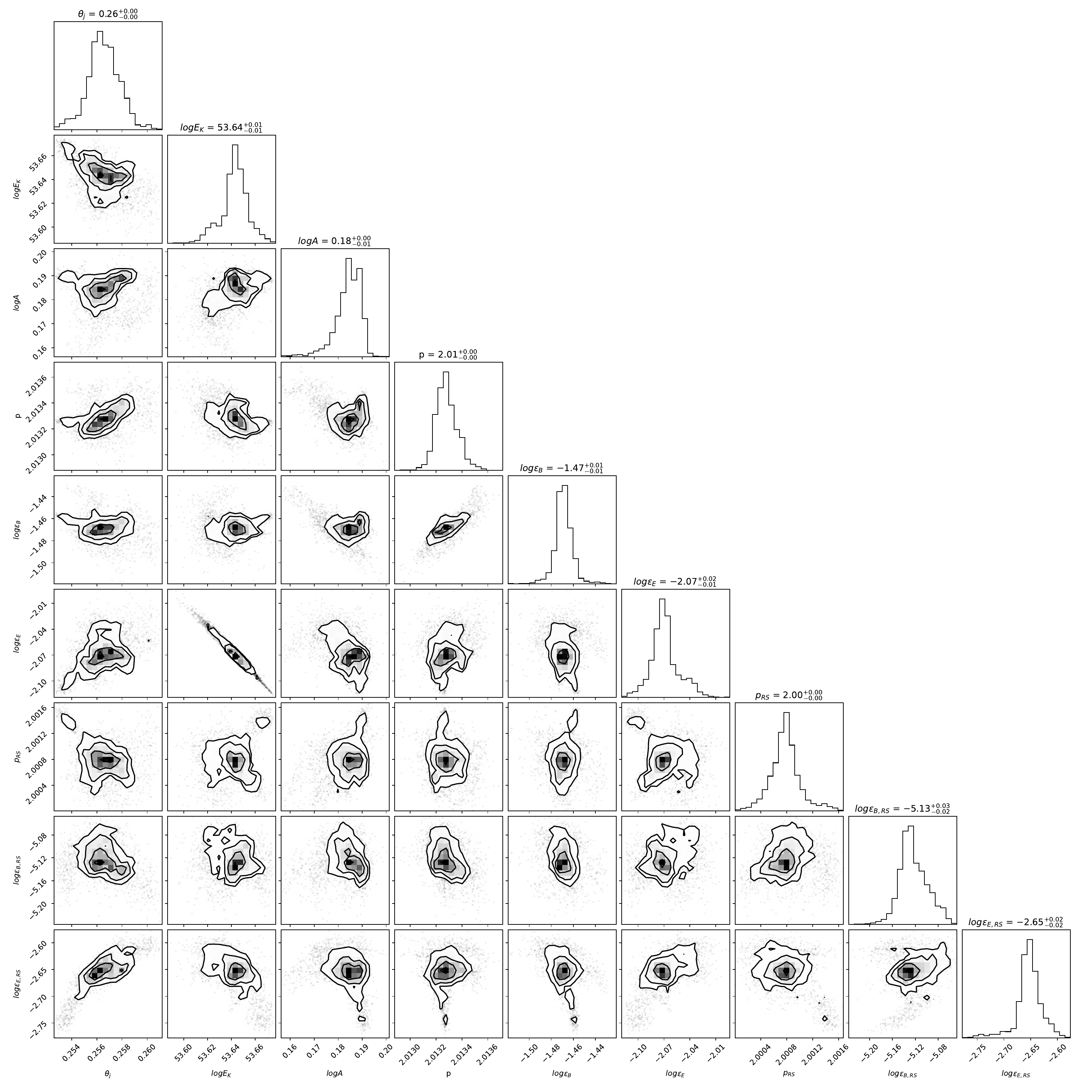}
      \caption{Corner plots for the wind medium fit.}
         \label{Fig:wind_fsrs}
   \end{figure*}

\begin{table*} %[!h]
\centering
\caption{Table of used standards in the field of GRB 250129A. Magnitudes were converted from SDSS system with the Lupton 2005 equations.}
 \begin{tabular}{cccccc}
     \hline\hline
\# & SDSS object          & B                  & V                  &  R                 & I                \\
    \hline
 1 & J131443.96+050109.9  & $18.390 \pm 0.053$ & $17.856 \pm 0.008$ & $17.544 \pm 0.010$ & $17.201 \pm 0.012$ \\
 2 & J131437.94+050129.0  & $18.549 \pm 0.060$ & $17.541 \pm 0.008$ & $16.939 \pm 0.009$ & $16.398 \pm 0.009$ \\
 3 & J131436.43+050122.6  & $18.988 \pm 0.074$ & $17.638 \pm 0.008$ & $16.829 \pm 0.009$ & $16.140 \pm 0.009$ \\
 4 & J131433.38+050141.1  & $19.471 \pm 0.094$ & $18.015 \pm 0.009$ & $16.973 \pm 0.009$ & $15.732 \pm 0.009$ \\
 5 & J131434.61+050205.6  & $17.140 \pm 0.051$ & $16.487 \pm 0.007$ & $16.097 \pm 0.009$ & $15.696 \pm 0.008$ \\
 6 & J131447.60+050417.4  & $16.325 \pm 0.050$ & $15.713 \pm 0.006$ & $15.345 \pm 0.008$ & $14.963 \pm 0.008$ \\
 7 & J131454.43+045903.3  & $16.346 \pm 0.051$ & $15.717 \pm 0.007$ & $15.349 \pm 0.008$ & $14.988 \pm 0.008$ \\
 8 & J131450.97+045955.8  & $18.415 \pm 0.054$ & $17.799 \pm 0.008$ & $17.434 \pm 0.009$ & $17.047 \pm 0.011$ \\
 9 & J131443.99+045952.7  & $20.229 \pm 0.205$ & $18.785 \pm 0.011$ & $17.860 \pm 0.010$ & $16.946 \pm 0.011$ \\
10 & J131446.62+050251.1  & $19.116 \pm 0.057$ & $18.654 \pm 0.010$ & $18.377 \pm 0.012$ & $18.043 \pm 0.019$ \\
11 & J131425.61+050400.3  & $18.127 \pm 0.012$ & $17.171 \pm 0.007$ & $16.606 \pm 0.009$ & $16.109 \pm 0.009$ \\
    \hline\hline       
\end{tabular}
\label{standards}
\end{table*}

\begin{table*}
\centering
\caption{Optical photometry of GRB~250129A using multiple telescopes over the globe. The magnitudes are corrected for Galactic extinction is E($B-V$) = $0.039$ mag, or $A_V$ = 0.120 mag \citep{Schlafly2011}.}
 \begin{tabular}{lcccr}
   \hline
$UT_{start}$                    & Time since burst    &  filter &  Magnitude   &  Instrument     \\
                & (hours) &   &   mag & \\
   \hline

2460705.58815 & 21.36 & B & $19.93\pm 0.12$ & LCO/1.0-m Sinistro  \\
2460705.59446 & 21.51 & B & $19.84 \pm 0.01$ & LCO/1.0-m Sinistro  \\
2460705.60864 & 21.85 & B & $19.85 \pm 0.02$ & LCO/1.0-m Sinistro  \\
2460706.23123 & 36.80 & B & $20.44 \pm 0.11$ & LCO/1.0-m Sinistro  \\
2460706.82903 & 51.14 & B & $20.91 \pm 0.02$ & LCO/1.0-m Sinistro  \\
2460707.44917 &   66.03    & B & $20.90\pm 0.07$ & DFOT/ CCD-camera \\
2460707.50393  &  66.83 &  B      &  $20.98 \pm 0.03$ &  Zeiss-1000/CCD-phot. \\
2460709.58604 & 117.31 & B & $22.40 \pm 0.09$ & LCO/1.0-m Sinistro  \\

\hline

2460704.76597 & 1.63  & V & $17.34 \pm 0.01$ & LCO/0.4-m SCICAM QHY600 \\
2460704.77642 & 1.88  & V & $17.41 \pm 0.01$ & LCO/0.4-m SCICAM QHY600 \\
2460705.08089 & 9.18  & V & $17.93 \pm 0.04$ & LCO/0.4-m SCICAM QHY600\\
2460705.56264 & 20.75 & V & $19.39 \pm 0.02$ & LCO/1.0-m Sinistro \\
2460705.57450 & 21.03 & V & $19.38 \pm 0.02$ & LCO/1.0-m Sinistro \\
2460705.62376 & 22.21 & V & $19.26 \pm 0.08$ & LCO/1.0-m Sinistro \\
2460706.57684 & 45.09 & V & $20.19 \pm 0.12$ & LCO/1.0-m Sinistro \\
2460706.94115 & 53.83 & V & $20.67 \pm 0.04$ & LCO/1.0-m Sinistro \\
2460706.99403 & 55.10 & V & $21.03 \pm 0.28$ & LCO/1.0-m Sinistro \\
2460707.01144 & 55.52 & V & $20.75 \pm 0.11$ & LCO/1.0-m Sinistro \\
2460707.41065 &  65.10     & V & $20.55 \pm 0.04$ & DFOT/ CCD-camera\\
2460707.50740   &  66.92  &  V      &  $20.52 \pm 0.02$ &  Zeiss-1000/CCD-phot.\\ 
2460709.35739  &  111.82    & V & $21.77 \pm 0.07$ & DFOT/ CCD-camera \\
\hline  
2460706.45637   &  42.20                 &  R      &  $19.80 \pm 0.04$ &  DFOT/ CCD-camera\\
2460706.50164   &  43.29          &  Rc     &  $19.82 \pm 0.02$ &  Zeiss-1000/CCD-phot. \\
2460707.36277   &  63.95          &  R      &  $20.15 \pm 0.04$ &  DFOT/ CCD-camera\\
2460707.45391   &  66.64         &  Rc     &  $20.19 \pm 0.02$ &  Zeiss-1000/CCD-phot. \\
2460708.56484  &  92.81         &  Rc     &  $21.25 \pm 0.14$ &  Zeiss-1000/MAGIC    \\
2460709.42937   &  113.55       & R     &  $21.20 \pm 0.07$  & DFOT/ CCD-camera\\
2460711.38756   &   160.55       & R      &  $22.43 \pm 0.1$  & DFOT/ CCD-camera\\
2460734.47631  & 714.604          &  Rc     &  $>24.00$            &  Zeiss-1000/CCD-phot.\\
\hline

%              & 0.57 & r & $16.71 \pm 0.02$ \\
2460704.74678 & 1.17  & r & $17.05 \pm 0.01$ & LCO/0.4-m SCICAM QHY600 \\
%              & 1.32  & r & $16.72 \pm 0.02$ \\
2460704.75724 & 1.42  & r & $17.17 \pm 0.01$& LCO/0.4-m SCICAM QHY600 \\
%              & 1.89  & r & $17.27 \pm 0.02$ \\
2460704.76597 & 8.88  & r & $17.75 \pm	0.01$ & LCO/0.4-m SCICAM QHY600\\
%              & 18.83 & r & $19.29 \pm 0.08$ \\
2460705.50914 & 19.46 & r & $19.16 \pm 0.01$& LCO/1.0-m Sinistro \\ 
2460705.52102 & 19.75 & r & $19.17 \pm 0.01$& LCO/1.0-m Sinistro  \\
%              & 21.44 & r & $19.24 \pm 0.03$ \\  
%              & 24.10 & r & $18.91 \pm 0.03$ \\  
%              & 31.71 & r & $19.38 \pm 0.04$ \\  
%              & 32.43 & r & $19.7 \pm 0.20$ \\  
%              & 37.83 & r & $19.74 \pm 0.16$ \\
%              & 49.44 & r & $20.29 \pm 0.04$ \\
2460706.26228 &   37.54    & r & $19.87 \pm 0.23$ & DOT/ ADFOSC\\
2460706.57392 &  45.02     & r & $20.06 \pm 0.17$& LCO/1.0-m Sinistro  \\
2460706.91924 & 53.30 & r & $20.46 \pm 0.15$ & LCO/1.0-m Sinistro \\
%              & 60.00 & r & $21.8 \pm 0.11$ \\ 
%              & 70.32 & r & $20.4	\pm 0.05$ \\
2460707.83711 & 75.33 &	r & $20.52 \pm 0.02$& LCO/1.0-m Sinistro  \\
2460709.58845 & 117.37& r & $21.87 \pm 0.07$ & LCO/1.0-m Sinistro \\
%              & 118.32& r & $22.10 \pm 0.15$ \\
%              & 122.83& r & $22.04 \pm 0.21$ \\
\hline
2460707.38832 &   64.57               &  I      &  $19.74 \pm 0.05$ & DFOT/ CCD-camera\\
2460707.45781 &  66.736          &  Ic     &  $19.73 \pm 0.05$ &  Zeiss-1000/CCD-phot. \\
\hline
2460706.50921 &     43.47             & i       & $19.71 \pm 0.06$ & DOT/ ADFOSC\\
2460708.51127 &    91.52              & i       & $20.90 \pm 0.11$ & DOT/ ADFOSC\\

     \hline
\end{tabular}
\label{tab:photometry}
\end{table*}

%%%%%%%%%%%%%%%%%%%%%%%%%%%%%%%%%%%%%%%%%%%%%%%%%%

% Don't change these lines
\bsp	% typesetting comment
\label{lastpage}
\end{document}